# Tungsten Boride: a 2D Multiple Dirac Semimetal for Hydrogen Evolution Reaction

Aizhu Wang[1,2], Lei Shen[3], Mingwen Zhao[4]*, Junru Wang[4], Weifeng Li[4], Weijia Zhou[1], Yuanping Feng[5]*, and Hong Liu[1,6]*

[1]*Institute for Advanced Interdisciplinary Research, University of Jinan, Jinan, Shandong, 250022, China*

[2]*Department of Electrical and Computer Engineering and Department of Physics, National University of Singapore, Singapore, 117579, Singapore*

[3]*Department of Mechanical Engineering, Engineering Science Programme, Faculty of Engineering, National University of Singapore, Singapore, 117575, Singapore*

[4]*School of Physics and State Key Laboratory of Crystal Materials, Shandong University, Jinan 250100, Shandong, China*

[5]*Department of Physics & Centre for Advanced Two-dimensional Materials, National University of Singapore, Singapore, 117542, Singapore*

[6]*State Key Laboratory of Crystal Materials, Shandong University, Jinan, 250100, Shandong, China*

*Correspondences:
zmw@sdu.edu.cn (M. Z.)
phyfyp@nus.edu.sg (Y. F.)
hongliu@sdu.edu.cn (H. L.)

**TOC**

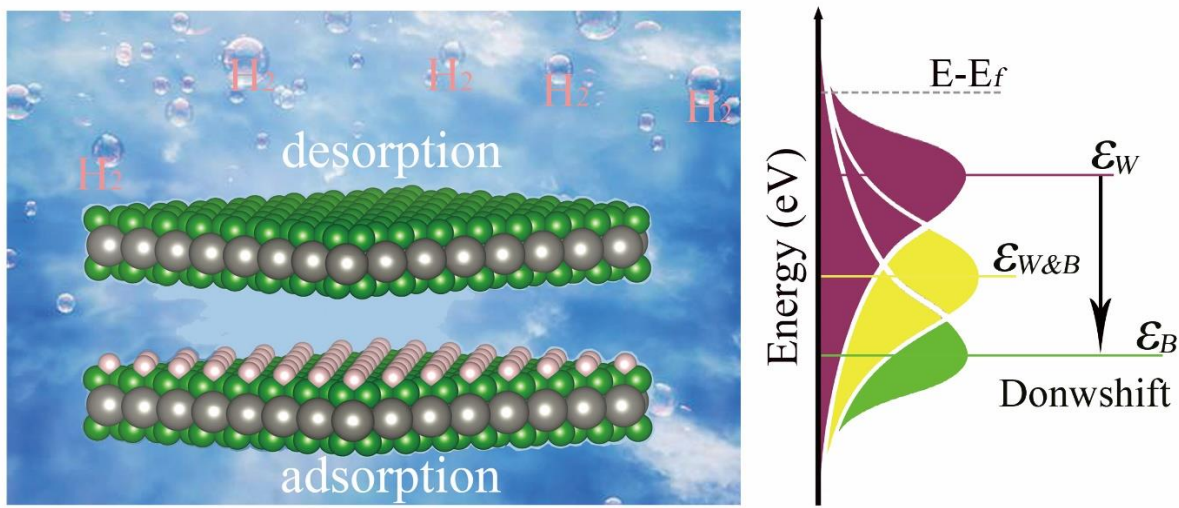

**HIGHLIGHTS**

2D $WB_4$ is first theoretical studied as an efficient electrocatalyst for hydrogen evolution reaction.

The 2D "sandwich" B-W-B lattice intrinsically possesses multiple Dirac states for facilitating the transmission of electrons.

The optimized catalyst shows an ultralow overpotential to realize its ideal catalytic characteristics.

The $d$ band center regulated by the $p$-orbitals of borophene subunits is the main driving force for achieving higher activity.

High Pt-like HER activity of the host is essential to offer a guiding principle for discovering optimal noble-metal-free catalysts.

**SUMMARY**

Here, we propose a two-dimensional (2D) tungsten boride (WB$_4$) lattice, with the Gibbs free energy for the adsorption of atomic hydrogen ($\Delta G_H$) tending to be the ideal value (0 eV) at 3% strained state, to host a better hydrogen evolution reaction (HER) activity. Based on first-principles calculations, it is demonstrated that the multiple *d-p-π* and *d-p-σ* Dirac conjugations of WB$_4$ lattice ensures its excellent electronic transport characteristics. Meanwhile, coupling with the *d*-orbitals of W, the *p*-orbitals of borophene subunits in WB$_4$ lattice can modulate the *d* band center to get a good HER performance. Our results not only provide a versatile platform for hosting multiple Dirac semimetal states with a "sandwich" configuration, but also offer a guiding principle for discovering the relationship between intrinsic properties of the active center and the catalytic activity of metal layer from the emerging field of 2D noble-metal-free lattices.

**Context & Scale**

Efficient hydrogen production ($H_2$) from water could potentially lead to a clean and renewable energy system. Electrochemical water splitting to produce $H_2$ offers a reliable and sustainable solution for this purpose. The catalysts are important components of water electrolysis cells and largely govern their performance. Considering the high price and limited world-wide supply of Pt-based HER catalysts, the development of non-noble-metal and high activity hydrogen-producing catalysts plays a central role in the current research progress.

In this work, we propose a 2D $WB_4$ lattice, with the $\Delta G_H$ tending to be the ideal value (0 eV) at 3% strained state, host a better HER activity than that in Pt ($\Delta G_H \sim -0.05$ eV). Based on first-principles calculations, we present a systematic theoretical study for the $WB_4$ lattice with special emphasis on the configuration design and electronic structure, and find that the $WB_4$ lattice spawns multiple Dirac cones around the Fermi level with considerable fermi velocities to transfer electrons in all directions throughout its structure. Importantly, together with the *d*-orbitals of W, the *p*-orbitals of borophene subunits in $WB_4$ lattice can modulate the *d* band center to get a good HER performance by regulating external stress. This work provides new insight into the rational design and manipulation of 2D catalysts toward electrochemical hydrogen production and other electroreduction reactions with high selectivity.

**INTRODUCTION**

When powered by renewable energy sources, the generation of hydrogen from the electrolysis of water is a means to produce a high-energy density, mobile energy carrier without any associated carbon dioxide emissions [1]. At low temperatures, this process can take place in a series of electrochemical devices, ranging from high-current-density polymer electrolyte membrane electrolyzers [2-4] to low-current-density, solar-driven photoelectrochemical cells [5-7]. For all those water-splitting applications, the choice of the hydrogen evolution reaction (HER) catalyst employed at the cathode can have a profound influence on the cost, lifetime, and efficiency of the device. Pt is a very active and commonly used HER catalyst, but its high price and limited world-wide supply [1, 8] make its use a barrier to mass production of $H_2$ by water electrolysis.

The effective approach to overcome the challenges associated with Pt HER catalysts is to increase the surface to bulk atomic ratio of Pt, thus allowing for a lower metal loading to be used without compromising electrolysis efficiency [1]. Many efforts have really been devoted both experimentally and theoretically in the past few years. For example, chemical treatments [9-11], such as surface modification, alloying, heteroatom doping and so on. Another approach that works is completely replacing Pt with alternative non-Pt catalysts. Such as transition metal alloys, sulfides, carbides, nitrides, and borides, *etc*. [12-16]. Hence, a prerequisite is gaining detailed knowledge of structure-activity correlations for electrode materials to better understand the activity trends and then rapidly discovering new advanced catalysts for HER.

Sitting around carbon in the periodic table, boron is a magic element in the sense that it not only can bond both covalently and ionically [17], but it is capable of forming a great variety of pure allotropes ranging from zero-dimensional clusters to three-dimensional (3D) bulk materials [18-21], suggesting that boron share many of the characteristic of carbon. Advanced research demonstrated that several 3D boride compounds consists of two-dimensional (2D) boron layers as subunits (known as borophene subunits), have recently emerged as a potential new frontier for searching better HER catalysts [22-24], which can be contributed as the strongly bound of borophene subunits within the solid in contrast to the weakly bound graphene layer in graphite. However, the research on the HER activity of 2D borophene subunits is still a blank.

Here, the 2D tungsten boride (WB$_4$) lattice (**Figure 1(a)**) is proposed, in which a triangular tungsten layer is sandwiched by two borophene subunits, forming a "sandwich" layer. Based on first-principles calculations, we find that the WB$_4$ lattice has superior stability and multiple Dirac cones due to the *d-p-π* and *d-p-σ* conjugations. Moreover, our study demonstrates a different principle to spur HER activity that can go beyond local site optimization by manipulating the *d* band center within a certain strain range, as evidenced by an optimum Gibbs free energy of 2×2 WB$_4$ nanosheet. Our results, here, not only provide a versatile platform for hosting multiple Dirac semimetal states with a "sandwich" configuration, but also offer a guiding principle for discovering the relationship between intrinsic properties of the active center and the catalytic activity of metal layer from the emerging field of noble-metal-free lattices.

## RESULTS AND DISCUSSION

**Characterization of WB$_4$**

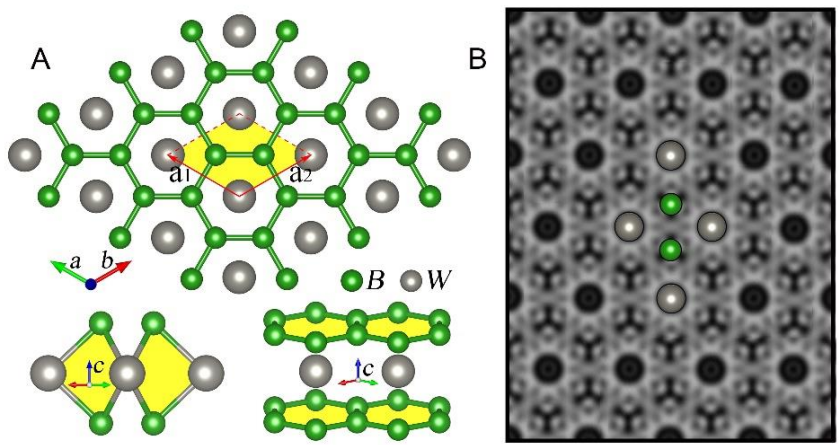

**Figure 1. Physical Characterization of WB$_4$**

(A) Schematic representation of the 2D honeycomb tungsten boride (WB$_4$). The unit cell is indicated by the yellow shaded area with the two basis vectors of **a$_1$** and **a$_2$**.
(B) Simulated STM image for monolayer WB$_4$.

Experimentally, the WB$_2$ is found in the hexagonal phase within P6/mmm space group with few freedom of atomic positions [25]. Only on this alone, WB$_4$ monolayer can be naturally produced through exfoliation of the nanosheet (Supplemental Information). **Figure 1A** represents the top view and side view, respectively, of the optimized configuration, where the triangle lattice W atoms are located above the center of the honeycomb graphene-like boron lattice from the top view, form a "sandwich". The lengths are 1.706 Å, 2.371 Å and 2.955 Å for covalent B-B, ionic W-B and metallic W-W bonds, respectively, and the optimized lattice constant is found to be 2.955 Å, matching with the trigonal W lattice in $WB_2$.

Then, the thermodynamic stability of the $WB_4$ framework is confirmed by the phonon spectrum calculated along the highly symmetric directions in BZ (**Figure S3**). Clearly, the phonon spectrum of $WB_4$ lattice is free from imaginary frequencies, implying the thermodynamic stability of $WB_4$ lattice. By comparing the cleavage energy, it is verified that $WB_4$ lattice is robust stable than $WB_2$ in the process of synthetic 2D tungsten boride lattice. So, it can be regarded as $WB_4$ multilayers, and a single layer of large dimensions is plausible in near future thanks to the well-developed preparative technique (Supplemental Information). To offer more reference information to the experimental observation, the simulated STM image of $WB_4$ monolayer is also displayed (**Figure 1B**).

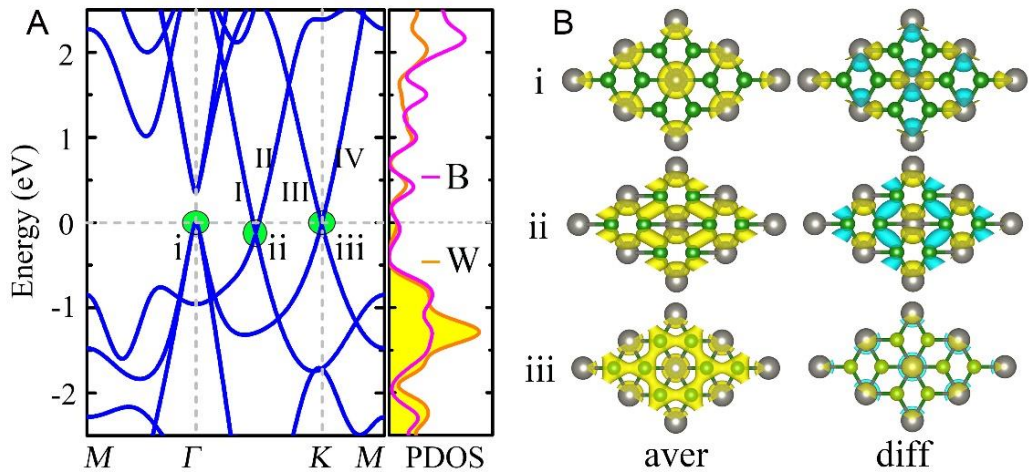

**Figure 2. Electrical Characterization of $WB_4$**

(A) Electronic band lines and the corresponding electronic density of states of $WB_4$ in proximity of the Fermi level (set to zero).

(B) The isosurfaces of the Kohn-Sham wave functions of degenerate points listed in (A). Different colors represent different bands in diff.

## Excellent Electronical Transport Property

The electronic properties including the band structure and corresponding projected electronic density of states are presented (**Figure 2A**). Based on density-functional theory (DFT) calculations, two Dirac cones exist closely to the Fermi level ($E_f$), exhibiting the characteristic of semimetals. By plotting the orbital-resolved band structures as well as the charge density distributions, it is clearly seen that the Dirac bands are composed mainly of the $p$ states of B atoms and the $d$ states of W atoms. More specifically, bands I and II are from the coupling of $p_{x/y/z}$ orbitals of B atoms and the $d_{z2}/d_{x2-y2}/d_{xy}$ orbitals of W atoms, meanwhile, bands III and IV are mainly derived from the coupling of $p_z$ orbital of B and W atoms, together with the $d_{xz}/d_{yz}$ orbitals of W atom (**Figure S4** and **Figure S5**).

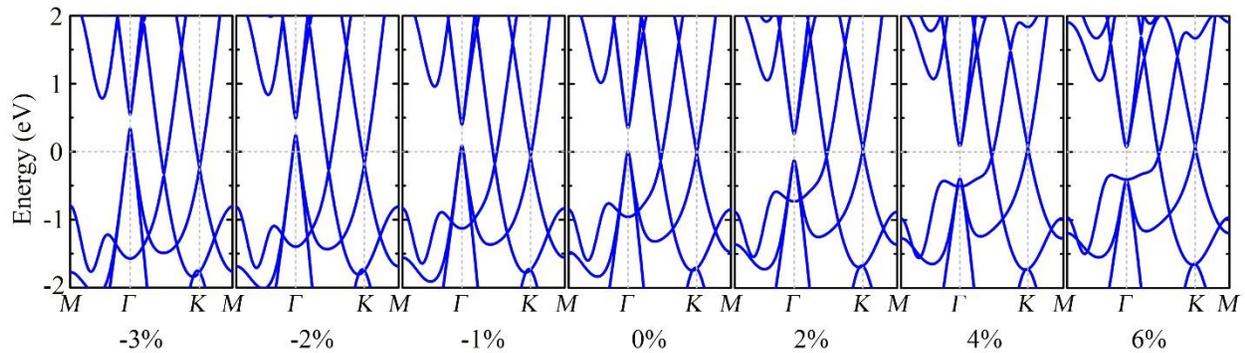

**Figure 3**. **Strain Response of the Semimetal State in WB$_4$**

Electronic band structure evolution of the 2D WB$_4$ lattice with external strain along the $xy$-direction. The energy at the Fermi level was set to zero. Here, ''-'' represents the condition of compressional strain.

Apart from the equalitarian state, we also investigate the robustness of semimetal states under finite biaxial strain. **Figure 3** shows that the semimetal feature is robust against the stretching along the $xy$-direction, but gradually converts into metal characteristic with abundant states around $E_f$ under the compressional condition, suggesting a different strain response of WB$_4$. The Fermi velocity ($v_f$) of the WB$_4$ lattice was calculated by fitting those Dirac bands and achieved extreme values at the equalitarian state, $0.72 \times 10^6$ m/s for band I, $0.54 \times 10^6$ m/s for band II and $0.61 \times 10^6$ m/s for bands III and IV (Supplemental Information). The multiple Dirac states (different location, origin and anisotropies/isotropies), together with their considerable fermi velocities, should enable

WB$_4$ monolayer to transfer electrons in all directions throughout its structure, suggesting its excellent electronic transport property.

**Hydrogen Evolution Reaction Activity**

The metallic property is favorable for the electron injection from a cathode to the catalyst surface, where intermediate protons induced by water dissociation in alkaline solutions are reduced and adsorbed on the catalyst-covered cathode [26]. Hence, we study electrocatalysis applications of the Dirac semimetal lattice with high carrier mobility. Both experimentally and theoretically, the Volmer step is recognized as the rate-determining step for HER [27-29], so, the Gibbs free energy of the adsorption atomic hydrogen ($\Delta G_H$) is the key operator to describe the HER activity [30]. Here, we adopt a (2×2) supercell of WB$_4$ lattice to study its catalytic activity and find that the bridge site is the most stable adsorption site (**Figure 4A** and Supplemental Information). Therefore, we will only focus on the bridge site hereafter.

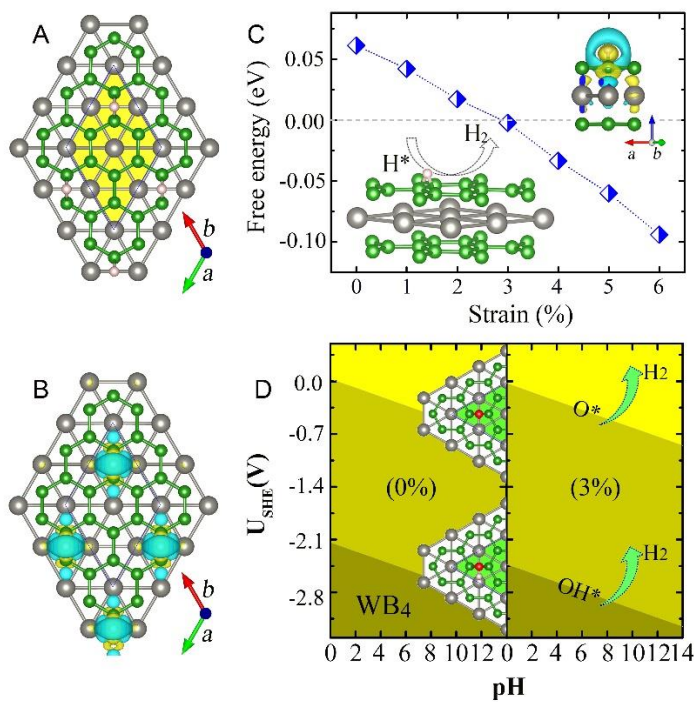

**Figure 4. Identification of Active Sites and HER Performance**
(A and B) Schematic representation of (2×2) WB$_4$ 2D supercell with a hydrogen adsorption (A) and the corresponding charge density difference before and after the adsorption (B). The unit cell is indicated by the yellow shaded area.

(C) Gibbs free energy diagram of HER in strained states. Blue and yellow represent charge loss and charge accumulation, respectively. The isosurface value is set to 0.002Å$^{-3}$.

(D) Surface pourbaix diagrams. The most thermodynamically stable states of the surface under relevant $U_{SHE}$ conditions and pH values are labeled by the terminations. The optimized structures of hydroxyl-terminated WB$_4$ and oxygen-terminated WB$_4$ were listed in the insert figure. Oxygen is represented by the red ball.

In order to elucidate the adsorption mechanism, we further analyze the charge transfer between with and without the hydrogen adsorption, which is defined as $\Delta Q = Q(sub+H)-Q(sub)-Q(H)$, where Q(sub+H), Q(sub) and Q(H) are the charge densities of the corresponding WB$_4$ systems with a hydrogen adsorption, the clean corresponding WB$_4$ systems and a single hydrogen atom, respectively. As **Figure 4B** shown, charges can transfer between B and H atoms as well as the nearest W atoms, providing a possible platform for catalytic reactions. Formally, the charges are mainly contributed by *p*-orbitals of B and *d*-orbitals of W (**Figure S7** and **Figure S8**). Quantitatively, the Bader charge analysis reveals that the hydrogen has a net gain of 2.32$e$ from the surrounding B and W atoms, exhibiting an excellent potential performance for HER activity.

Based on the PBE functional, we obtained the Gibbs free energy $\Delta G_H = 0.06$ eV for the above WB$_4$ nanosheet, indicating that the surface bridge site is catalytically active for HER. Inspired by the experimental stress effect on catalytic activity [31, 32], we investigated the biaxial strain effect on the HER catalytic activity of WB$_4$ lattice, and the related Gibbs free energies are calculated (**Figure 4C**). Interestingly, the $\Delta G_H$ of WB$_4$ nanosheet responds differently to the above-mentioned Dirac semimetal states. For example, the stretch makes the WB$_4$ nanosheet approaching the ideal catalytic state ($\Delta G_H \sim 0.00$ eV) firstly, and then far away the state with further increased stretch, suggesting that WB$_4$ nanosheet has a good HER performance within a certain tensile strain.

**Controlled *d* band Mechanisms**

To analyze the origin of the catalytic activity evolution, we examined how the *d*-band center can be modulate with the change of strain in WB$_4$ nanosheet. Considering the orbital hybridization between W and W atoms (the distance ranges from 2.955 Å to 3.132 Å), the *d*-orbitals of four W atoms are plotting (**Figure 5)** and their *d*-band centers are marked by black arrows. According to

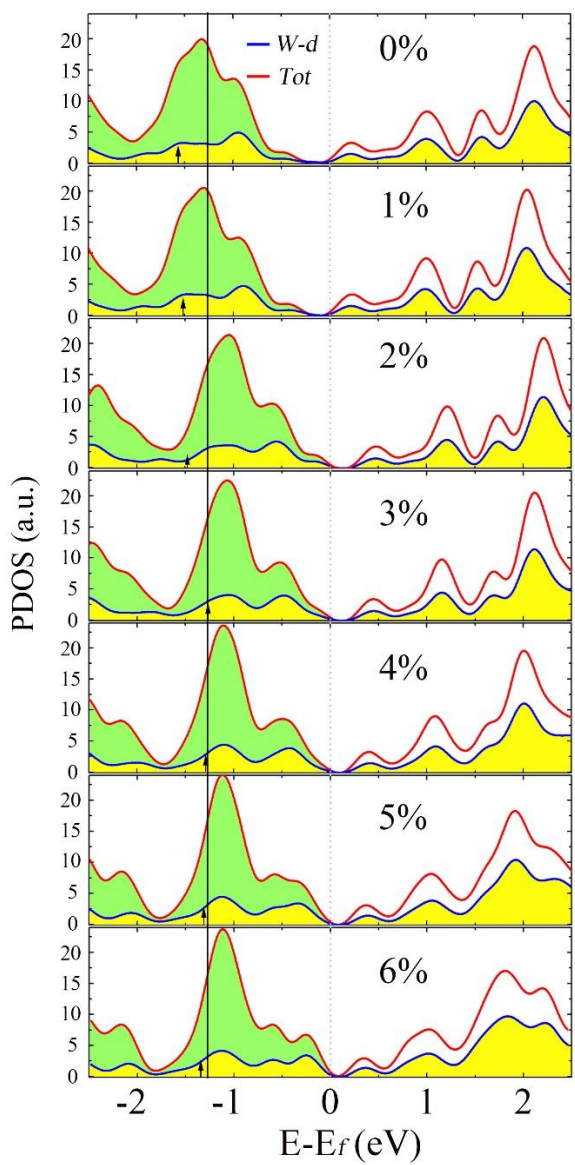

**Figure 5. Evolution of *d* Band Center**

The projected density of states for the *d*-orbitals of W in the strained states. The vertical black lines correspond to the location of strained states with 3% which has the highest catalytic activity. The energy at the Fermi level was set to zero.

the *d*-band theory, a lower *d*-band center results in weaker bonding between the catalyst and the adsorbate [33]. Taking the zero-point energy (ZPE) and entropy into account, the relative catalytic activity of WB$_4$ ($\Delta G_H$) has similar evolutionary relationship with the *d* band center (**Figure 4C**, **Figure 5** and **Figure S12**). The results can be attributed to the existence of B intermediate, which are modulating the combination of W with H to make a mild catalytic reaction by changing the coupling interactions between *d*-orbitals of W and *p*-orbitals of B. The coupling interactions makes the *d*-orbitals center of W shifting to a more negative value, resulting in a better catalytic activity. Additionally, the desorption barrier ($\Delta G_H \sim -0.06$ eV) of 5% strained state equals the adsorption barrier ($\Delta G_H$ ~ 0.06 eV) of 0% equalitarian state, so we can get the maximum adsorption and desorption by manipulating a small external stress to get an ideal catalytic state.

Except the modulated *d*-band center, the excellent conductivity and Fermi velocity, induced by the effect of Dirac cone with *d-p-π* and *d-p-σ* conjugations, also can guarantee fast electron transfer on the process of electrochemical reactions. Therefore, our results not only provide a guiding principle for the discovery of Dirac semimetal catalyst with noble-metal-free "sandwich" configuration, but also offer a universal description of the relationship between intrinsic properties of the active center and the catalytic activity of metal layer, which are in line with new research frontier in single-atom electrocatalysts [34-36].

**High Thermal Stability**

The surface Pourbaix diagrams of the equalitarian (0%) and strained (3%) states are constructed by plotting the thermodynamically most stable surface state under the relevant $U_{SHE}$ and pH values (**Figure 4D**). In an acidic solution (pH = 0), negative $U_{SHE}$ values (reducing environment) as low as -2.14 and -2.43 V are required to protect the equalitarian state and strained state, respectively, from oxidation by $H_2O$. A more negative potential is necessary to protect the bare $WB_4$ lattice at high pH values with a slope of -0.059 V/pH. When the potential is above the cathodic protection potential of the 2D $WB_4$, water starts to oxidize and the $WB_4$ is covered by one hydroxyl (OH*). As the potential increases, the hydroxyls on $WB_4$ will be oxidized, and the most stable O*-terminated $WB_4$ will be formed. The lowest $U_{SHE}$ values for 2D $WB_4$ with one O* termination are 0.03 V (0%) and -0.03 (3%) in an acidic solution. Therefore, under standard conditions, the strained state prefers to be terminated by O*, exhibits a better HER performance fitting with $\Delta G_H$ descriptor (Supplemental Information).

**Hydrogen Coverage Effect in $WB_4$**

Finally, we explore the hydrogen coverage effect on the HER catalytic performance of the 0% and 3% states. We define the coverage of hydrogen as the fraction of a monolayer with respect to the number of available B-B bridge sites (12 sites) in the basal plane. Here, we consider one ($\sigma = 1/12$), two ($\sigma = 1/6$), three ($\sigma = 1/4$) and four ($\sigma = 1/3$) hydrogen adsorption in 2×2 $WB_4$ supercell and calculate Gibbs free energies as summarized in **Figure 6** and Supplemental Information. As a benchmark, we firstly calculate the HER performance of Pt with $\Delta G_H$ = -0.05 eV, fitting well with other previous works [30, 34]. Interestingly, the variation of the Gibbs free energy is obvious with the increase of H coverage, indicating that $\Delta G_H$ is sensitive to the hydrogen coverage. Typically, the 3% strained state at $\sigma =1/3$ has a free energy of 0.02 eV, whose catalytic performance even exceeds that of Pt. Hence, considering the Pt-like activity for HER of $WB_4$, a series of the 2D noble-metal-free "sandwich" lattices should be promising materials to replace Pt or Pt compound in other electrochemical and photoelectrochemical applications.

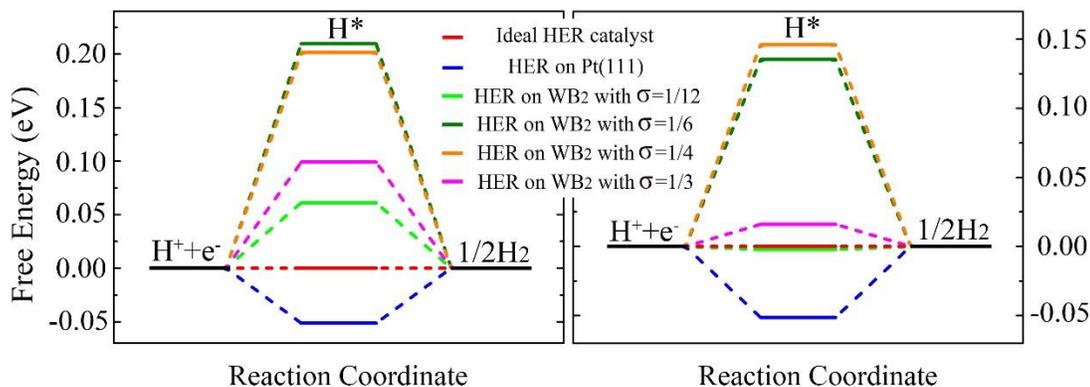

**Figure 6. Evolution of $\Delta G_H$ in Different Coverages**

Gibbs free energy diagram of HER under different coverage at equilibrium potential. As a benchmark, the Gibbs free energy for Pt was calculated by the same strategy and listed here.

**EXPERIMENTAL PROCEDURES**

**Computational Details of DFT Calculations**

The first-principles calculations within density-functional theory (DFT) were implemented by the Vienna *ab-initio* Simulation Package (VASP) [37]. A generalized gradient approximation (GGA) in the form of Perdew-Burke-Ernzerhof (PBE) [38] was adopted to describe the electron-electron interactions. The energy cut-off employed for plane-wave expansion of electron wavefunctions was set to 520 eV and the electron-ion interactions were treated using projector-augmented-wave (PAW) potentials [39]. The DFT-D3 method was applied to include the long-range van der Waals interaction for the adsorption on $WB_4$ lattice [40]. The system was modeled by unit cells repeated periodically on the *x-y* plane, while a vacuum region of 15 Å was applied along the *z*-direction to avoid mirror interaction between neighboring images. The Brillouin zone (BZ) integration was sampled on a grid of 13×13×1 *k*-points for the unit cell and 7×7×1 *k*-points for the 2×2×1 supercell.

Structure optimization was carried out using a conjugate gradient (CG) method until the remaining force on each atom is less than 0.05 eV/Å.

**Gibbs Free Energy of Hydrogen Adsorption**

The adsorption Gibbs free-energy of hydrogen, $\Delta G_H$, is a good descriptor for hydrogen evolution [41, 42], which is defined as $\Delta G_H = \Delta E_H + \Delta E_{ZPS} - T\Delta S_H$. $\Delta E_H$ is the hydrogen adsorption energy, which can be obtained as following: $\Delta E_H = E_{nH^*} - E_{(n-1)H^*} - 1/2 E_{H_2}$, where * donates the catalyst activity site. $E_{nH^*}$, $E_{(n-1)H^*}$ and $E_{H_2}$ are total energies of catalyst with $n$ adsorbed hydrogen atoms, total energies of catalyst with $n$-1 adsorbed hydrogen atoms and gas $H_2$, respectively. $\Delta E_{ZPE}$ and $\Delta S_H$ stand for the change in the ZPE and the change in the entropy of hydrogen between the adsorbed state and the gas state. We found that $\Delta E_{ZPE}$ does not vary significantly with respect to hydrogen coverage because of the smaller entropy of hydrogen in the adsorbed state. $T\Delta S_H \approx TS_H^0 = -0.202$ eV at $T = 298$ K, where $S_H^0$ is the entropy in the gas phase [43]. The standard hydrogen electrode ($U_{SHE}$) was theoretically defined in solution [pH = 0, p($H_2$) = 1 bar]. The optimal HER activity can be achieved as $\Delta G_H$ goes to zero, where both hydrogen adsorption and the subsequent desorption can be facilitated.

## ACKNOWLEDGMENTS

M.W.Z. thanks the support from the National Natural Science Foundation of China (Nos. 21433006 and 11774201). Y. P. Feng acknowledges funding support from the A*STAR's Pharos Programme on Topological Insulators (Grant No. 1527400026). The computational resources were provided by the Centre for Advanced 2D Materials of the National University of Singapore in Singapore and the National Super Computing Centre in Jinan.

## DECLARATION OF INTERESTS

The authors declare no competing interests.

# Supporting Information for

# Tungsten Boride: a 2D Multiple Dirac Semimetal for Hydrogen Evolution Reaction

Aizhu Wang[1,2], Lei Shen[3], Mingwen Zhao[4]*, Junru Wang[4], Weifeng Li[4], Weijia Zhou[1], Yuanping Feng[5]*, and Hong Liu[1,6]*

[1]*Institute for Advanced Interdisciplinary Research, University of Jinan, Jinan, Shandong, 250022, China*

[2]*Department of Electrical and Computer Engineering and Department of Physics, National University of Singapore, Singapore, 117579, Singapore*

[3]*Department of Mechanical Engineering, Engineering Science Programme, Faculty of Engineering, National University of Singapore, Singapore, 117575, Singapore*

[4]*School of Physics and State Key Laboratory of Crystal Materials, Shandong University, Jinan 250100, Shandong, China*

[5]*Department of Physics & Centre for Advanced Two-dimensional Materials, National University of Singapore, Singapore, 117542, Singapore*

[6]*State Key Laboratory of Crystal Materials, Shandong University, Jinan, 250100, Shandong, China*

*Correspondences:

zmw@sdu.edu.cn (M. Z.)

phyfyp@nus.edu.sg (Y. F.)

hongliu@sdu.edu.cn (H. L.)

## Part I The band structures of tungsten diboride

Figure S1(a) gives the top view and side view, respectively, of the tungsten diboride configuration, where the triangle lattice W atoms are located above the center of the honeycomb graphene-like boron lattice from the top view, form a "sandwich". The bond lengths are 1.746 Å, 2.513 Å and 3.025 Å for B-B, W-B and W-W, respectively. Values of a = 3.025 and c = 3.329 is calculated for a hexagonal unit cell, fitting well with the experiment data (a = 3.020 and c = 3.050). Clearly, the bulk materials are metallic with several bands across the Fermi level, as shown in the electronic band structure (Figure S1(b)). This is in good agreement with good electron conductivity of the tungsten diboride found in experiments [1].

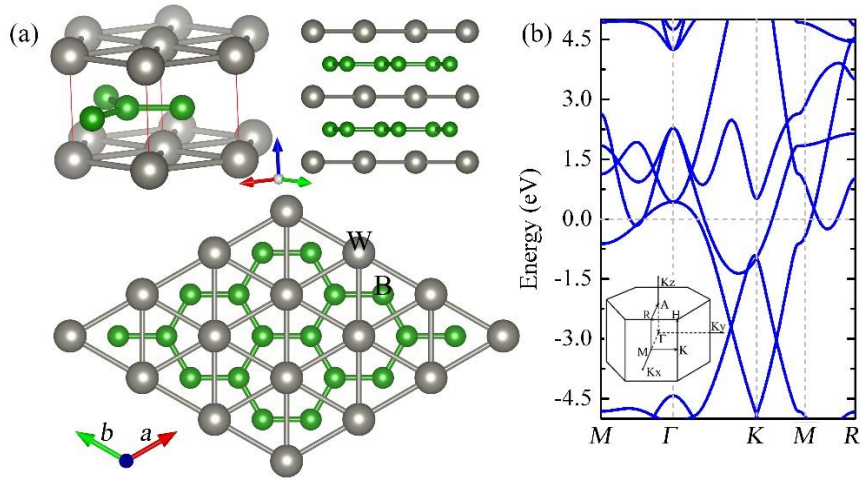

**Figure S1** (a) Schematic illustration of fabrication of $WB_2$ (P6/mmm, number 191) crystal. The unit cell is indicated by the red line. (b) The calculated band structures based on PBE calculations were also listed here. The energy at the Fermi level was set to zero.

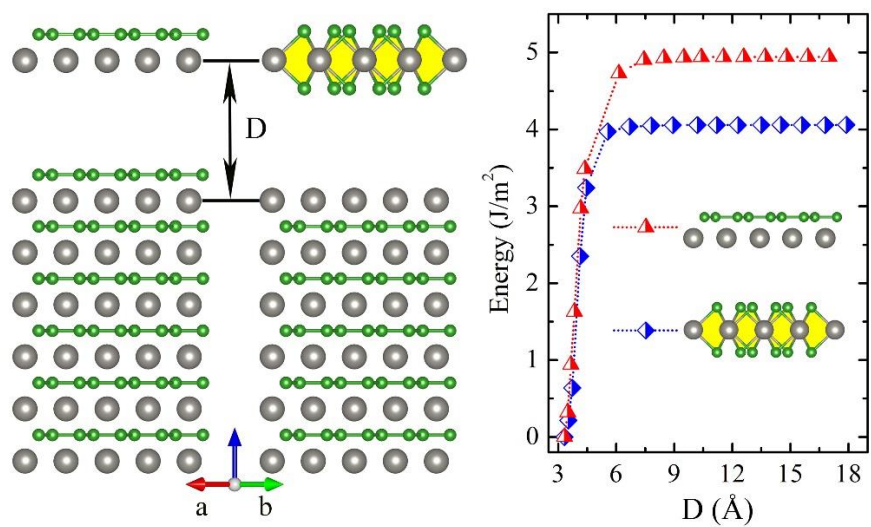

**Figure S2** (a) Schematic representation of the exfoliation process of WB$_2$ monolayer and WB$_4$ monolayer. (b) Energy increase $E$ as a function of interlayer distance (D).

Then, the possibility of producing WB$_4$ monolayer using a mechanical exfoliation strategy was confirmed (Figure 2S). The cleavage energy $E_{cl}$ is defined as the minimum energy required to exfoliate a monolayer from bulk. We used a seven-slab model to mimic a bulk material and calculated the energy increase as a WB$_4$ monolayer is exfoliated from the slab. A vacuum layer at least 15 Å was incorporated into the seven-layer slab to avoid the artificial interaction between two neighboring slabs. Figure S2 gives the variation of energy (and its derivative) as a function of the interlayer distance (D) between the top most monolayer and the remnant layers, which was fixed during the exfoliation process. The calculated cleavage energy $E_{cl}$ of WB$_4$ is about 4.06 J m$^{-2}$. The cleavage strength ($\chi$) was further obtained from the derivative of energy with respect to the distance, which is about 2.66 GPa. It is noteworthy that the calculated cleavage energy of WB$_4$ is smaller than that of WB$_2$ suggesting high plausibility to extract the WB$_4$ monolayer from the bulk in experiments.

## Part II The thermodynamic stability and electronical properties of WB$_4$

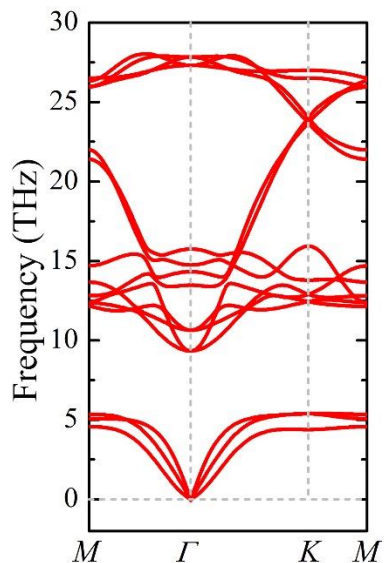

**Figure S3** (Color online) Phonon spectrum of WB$_4$ along the high symmetric points in BZ.

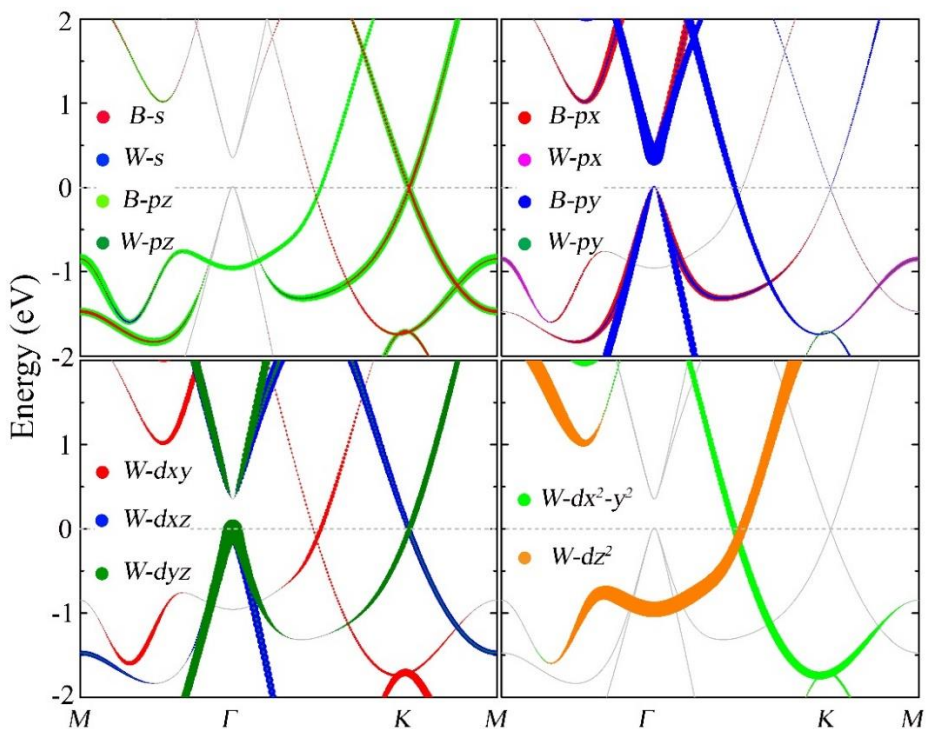

**Figure S4.** (Color online) Orbital-resolved band structures around Fermi level based on PBE calculations. The size of colorful dots is proportion to the contribution of the different orbitals on the wave function. The energy at the Fermi level was set to zero.

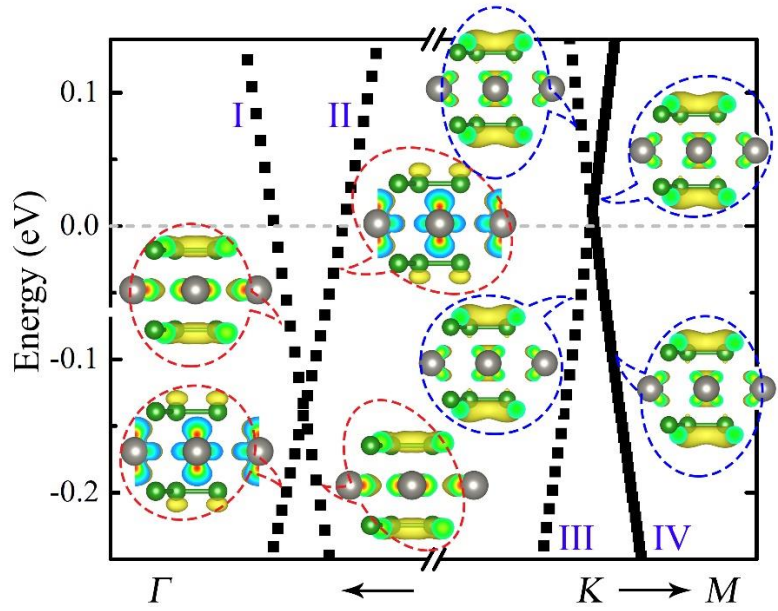

**Figure S5.** Charge density distributions near the Fermi level, both Dirac cone (cone ii and cone iii) are from *d* orbitals of W and *p* orbitals of B atoms. Cone iii, made up of bands III and IV, is isotropic with a high symmetry, while, cone ii, made up of bands I and II, is anisotropic with a lower symmetry. The energy at the Fermi level was set to zero.

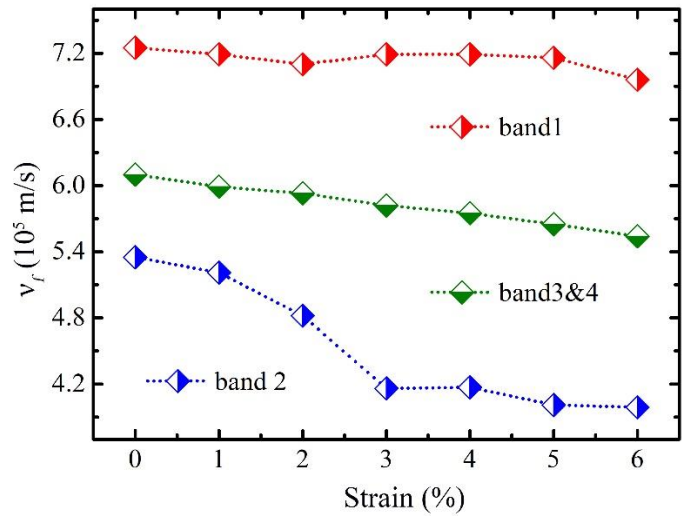

**Figure S6** Fermi velocity evolutions of Dirac bands in the 2D $WB_4$ lattice with external stretching strain along the *xy*-direction.

The charge carriers in these linear bands will behave as massless Dirac fermions. The Fermi velocity ($v_f$) of the WB$_4$ lattice can be tested by fitting those Dirac bands at $k = K + q$ to the expression of $v_f = E(q)/\hbar|q|$. The Fermi velocities are $0.72 \times 10^6$ m/s (band I), $0.54 \times 10^6$ m/s (band II) and $0.61 \times 10^6$ m/s (bands III and IV), respectively. These values are approximately 83.72%, 62.72% and 70.85% of that of graphene ($0.86 \times 10^6$ m/s) from the present calculations. Apart from the equalitarian state, we also investigated the robustness of semimetal states and Fermi velocities of the WB$_4$ monolayer under finite biaxial strain (-3% ~ 6%). Interestingly, the electronic structure of the WB$_4$ lattice responds differently to compressive (negative sign) and tensile (positive sign) strains as shown in **Figure 3**. The semimetal feature is robust against the stretching along the *xy*-direction, but gradually converts into metal characteristic with abundant states around $E_f$ under the compressional condition. As for those semimetal states, thanks for the configure symmetry and the out-of-plain coupling of orbitals ($p_z$ and $d_{xz}/d_{yz}$), the Dirac cone iii remains intact, but with a linear decreasing $v_f$ from $0.61 \times 10^6$ m/s to $0.55 \times 10^6$ m/s. Affected by $\Gamma$ point, the $v_f$ of band I almost keeps a constant, however, the value of band II has a sharp drop as shown in **Figure S6**.

**Part III The hydrogen evolution reaction calculations**

Here, we adopt a (2×2) WB$_4$ 2D supercell to study it's HER performance. On the basal plane of WB$_4$, there are three potential hydrogen adsorption sites, *i.e.*, W top site, B top site and the bridge site of B-B bond. The hydrogen adsorption energies on these sites are calculated and summarized in Table 1. The bridge site of B-B bond is the most stable adsorption site with ΔE$_H$ = -0.34 eV. The adsorption of hydrogen is much weak for the B top site ( -0.17 eV) and is unfavorable for the W top sites (positive values). Therefore, we will only focus on bridge site hereafter.

**Table S1**. Hydrogen adsorption energies on potential adsorption sites of WB$_4$ and the height of the hydrogen atom above the surface.

| Adsorption site | | Adsorption energy (eV) | Δz (Å) |
|---|---|---|---|
| W top sites | 1 | 5.56 | 0.27 |
|  | 2 | 1.51 | 0.67 |
|  | 3 | 1.51 | 0.65 |
| B top site | | -0.11 | 1.28 |
| Bridge site of B-B bond | | -0.31 | 1.01 |

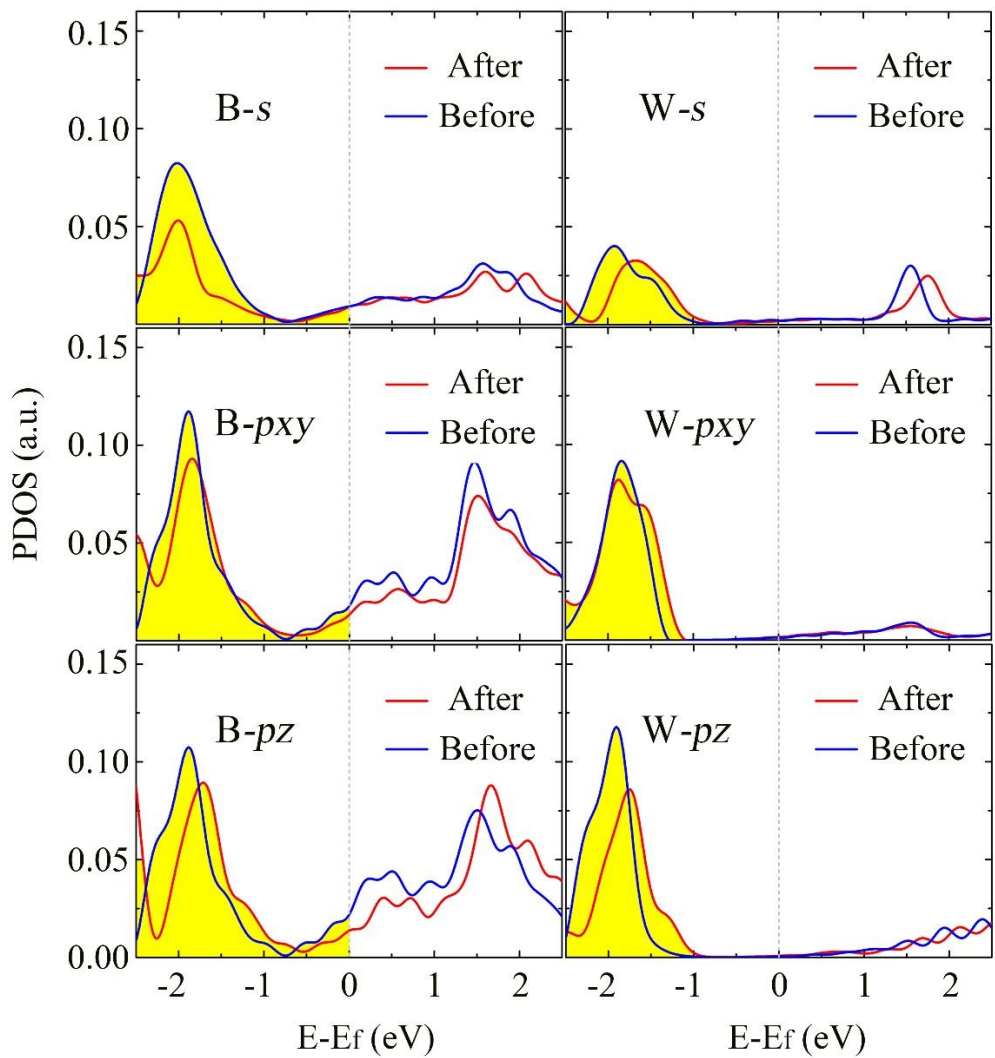

**Figure S7** (Color online) The projected density of states (PDOS) for the *s*- and *p*-orbitals of B and W atom under the condition with (after) and without (before) hydrogen adsorption. The energy at the Fermi level was set to zero.

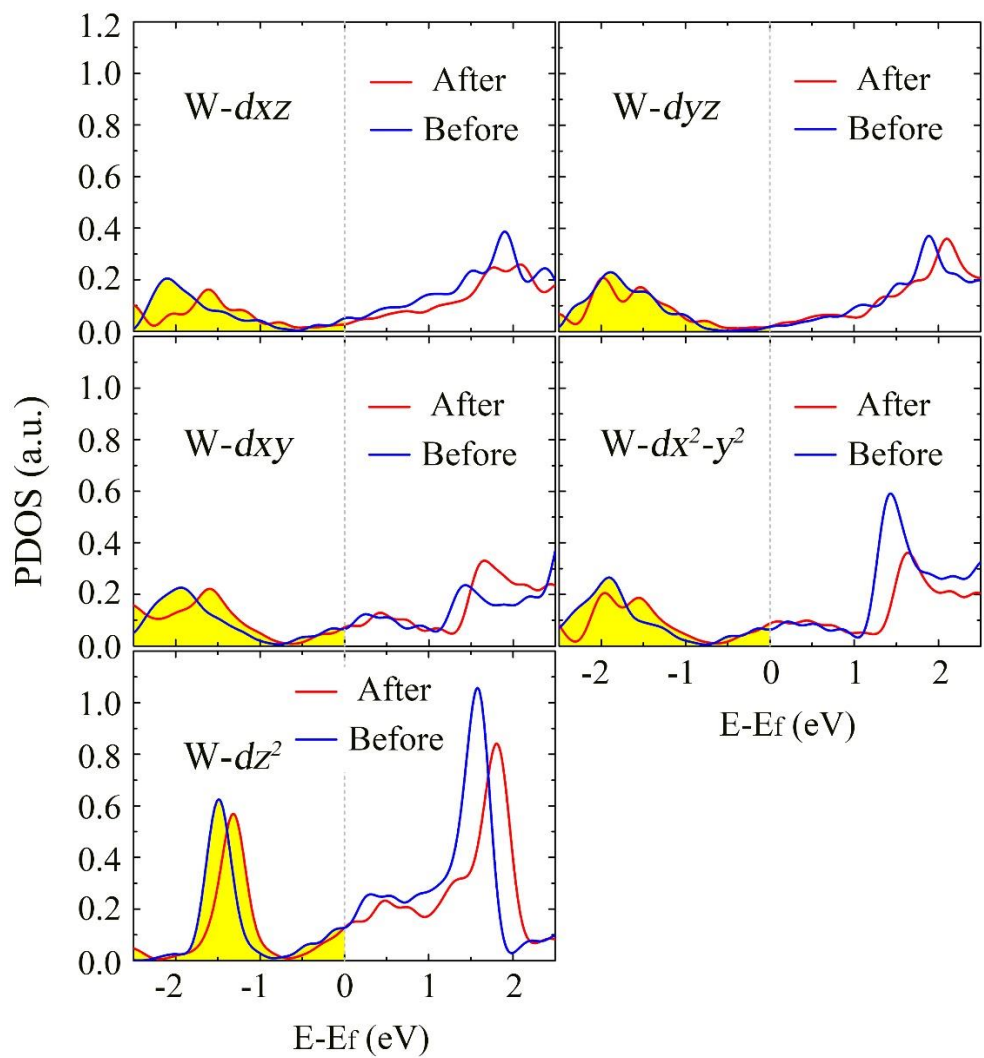

**Figure S8** (Color online) The projected density of states (PDOS) for the *d*-orbitals of W atom under the condition with (after) and without (before) hydrogen adsorption. The energy at the Fermi level was set to zero.

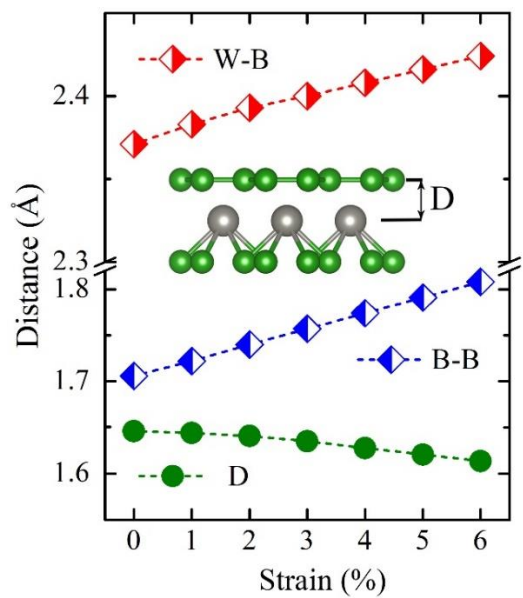

**Figure S9** (Color online) The evolutions of bonds. It can be clearly seen that the perpendicular distance between B and W decreases with the increase of lattice constant. And the hybridization between *p*-orbitals of B atoms and *d*-orbitals of W will change accordingly.

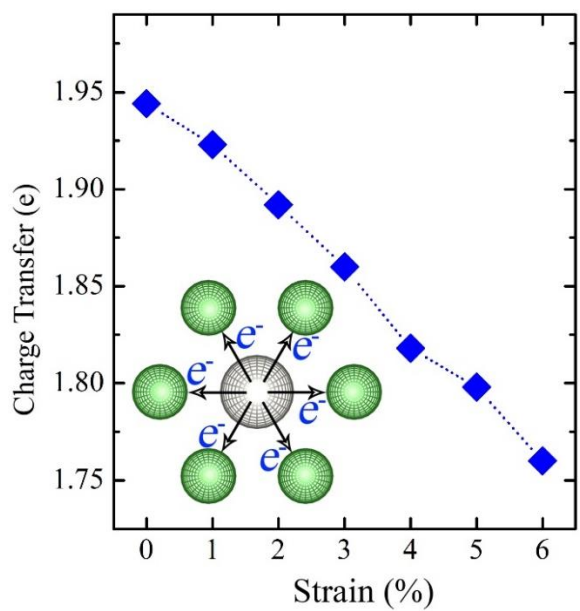

**Figure S10** (Color online) The evolutions of charge transfer from W atom to the surrounding B atoms in $WB_4$ nanosheet without hydrogen adsorption.

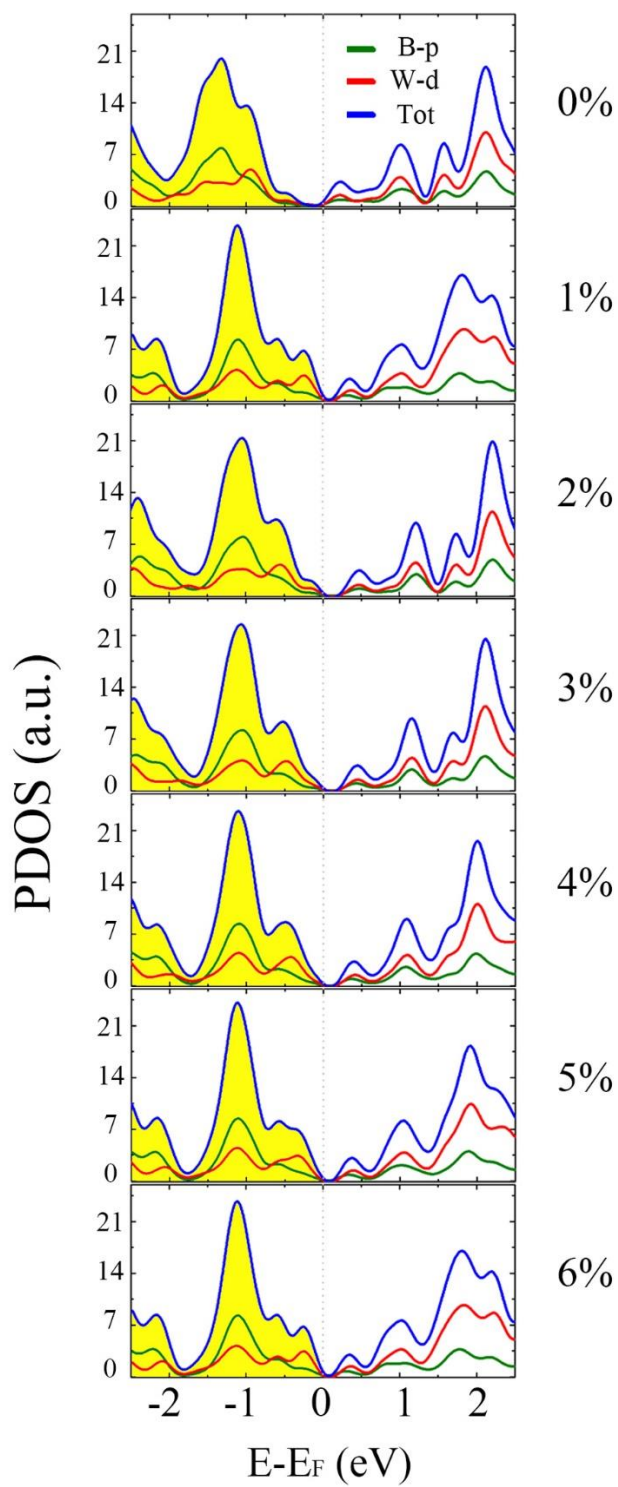

**Figure S11** (Color online) The evolutions of the projected density of states (PDOS) for B and W atoms. The energy at the Fermi level was set to zero.

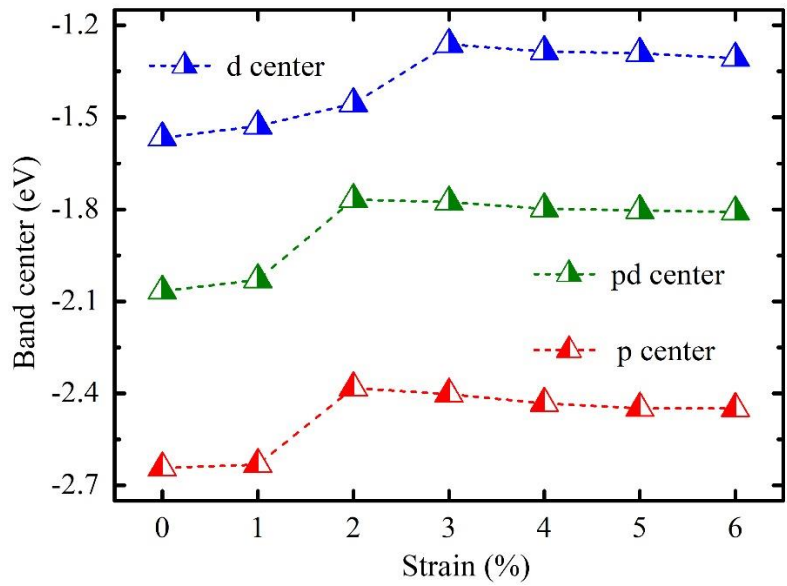

**Figure S12** (Color online) The evolutions of the band center. The energy at the Fermi level was set to zero. In the process of stretching, the *d* band center plays a leading role for the evolutions of catalyst activity.

The surface Pourbaix diagrams of the equalitarian (0%) and strained (3%) states are constructed by plotting the thermodynamically most stable surface state under the relevant $U_{SHE}$ and pH values. In our model, we assumed that the oxidation of water to OH* and O* on $WB_4$ lattice occurred through the following steps as suggested in ref [2]:

$$H_2O + * \rightarrow OH^* + H^+ + e^-$$
$$OH^* \rightarrow O^* + H^+ + e^- \quad (1)$$

Under standard conditions, the free energy of $H^+ + e^-$ is equal to 1/2 $H_2$. Therefor, the above equations can be rewritten as the following:

$$H_2O + * \rightarrow OH^* + \frac{1}{2}H_2 \quad \Delta G_1^0$$
$$OH^* \rightarrow O^* + \frac{1}{2}H_2 \quad \Delta G_2^0 \quad (2)$$

The Gibbs free energyies of $\Delta G_1^0$ and $\Delta G_2^0$ are obtained by

$$\Delta G^0 = \Delta E + \Delta E_{ZPE} - T\Delta S \quad (3)$$

where $\Delta E$ is the energy difference from with and without adsorbtion. The values from $E_{ZPE}$-$T\Delta S$ are calcualted on the basis of value from Table 1 of ref [3].

Equation (1) is deoendent on the pH and potential $U$ through the chemical potential of $H^+ + e^-$, while, equation (2) is not. To include the effects of pH and potential $U$, equation (3) can be rewritten as equation (4):

$$\Delta G_1 = \Delta G_1^0 - eU_{SHE} - k_bT \ln 10 \times pH$$
$$\Delta G_2 = \Delta G_2^0 - eU_{SHE} - k_bT \ln 10 \times pH \quad (4)$$

The free enrgy change of OH* and O* termination can be express by

$$\Delta G_{OH^*} = \Delta G_1$$
$$\Delta G_{O^*} = \Delta G_1 + \Delta G_2$$

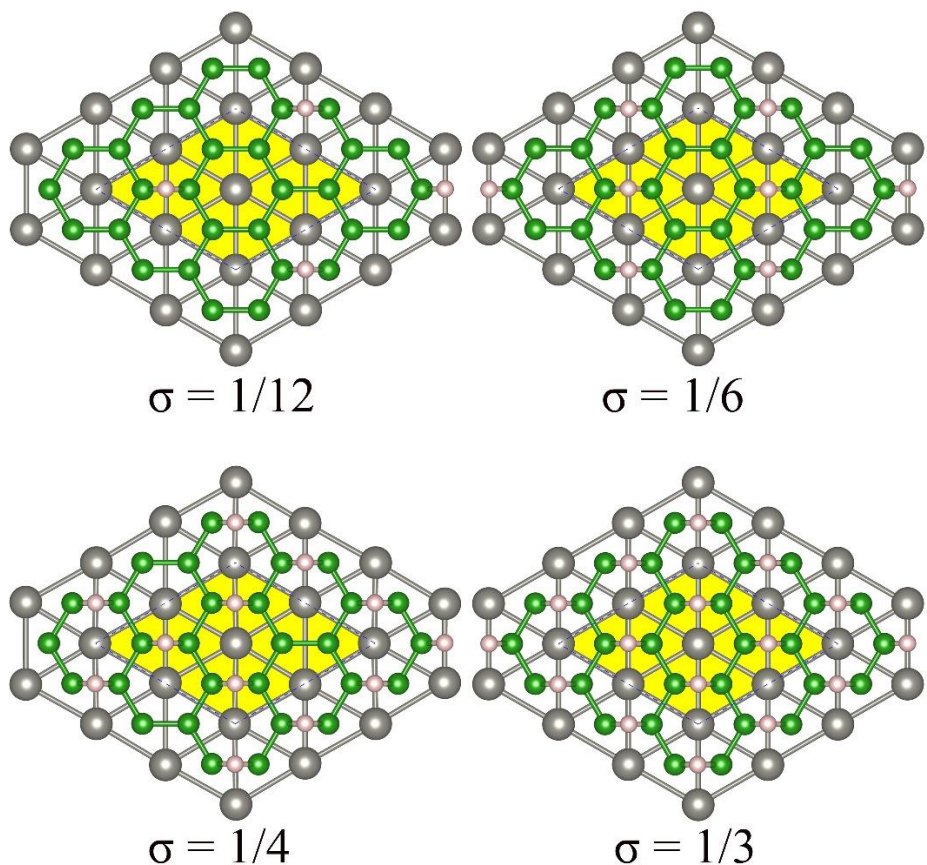

**Figure S13** Schematic illustration of stable adsorption states with different hydrogen covering. The unit cell is indicated by the yellow shaded area.